\documentclass[a4paper,10pt]{article}

\usepackage[english]{babel}
\usepackage[latin1]{inputenc}
\usepackage[T1]{fontenc}

\usepackage{amsmath}
\usepackage{amssymb}

\usepackage[authoryear,round,longnamesfirst]{natbib}

\newcommand{\roi}{\textsc{Roi}}

\title{Radial orbit instability~: review and perspectives}

\author{L.\ Maréchal\footnote{Electronic address~: lionel.marechal@ensta.fr}, %
	J.\ Perez\footnote{Electronic address~: jerome.perez@ensta.fr}\\%
	Laboratoire de Math\'{e}matiques Appliqu\'{e}es,\\%
	École Nationale Supérieure de Techniques Avancées,\\%
	32 Boulevard Victor, 75739 Paris cedex 15, France\\%
	Tel. 01 45 52 52 49, Fax 01 45 52 52 82}

\date\today

\begin{document}

\maketitle

\section*{Abstract}

This paper presents elements about the radial orbit
instability, which occurs in spherical self-gravitating systems with a strong
anisotropy in the radial velocity direction.
It contains an overview on the history of radial orbit instability.
We also present the symplectic method we use to explore stability of
equilibrium states, directly related to the dissipation induced instability
mechanism well known in theoretical mechanics and plasma physics.

\vspace{0.5ex}
\noindent \textbf{Keywords~:}
gravitation, instability, radial orbits, symplectic


\section{Introduction}

An interesting problem in Astrophysics is the study of $N$-body
self-gravitating systems with a lot of radial orbits, when most
particles have very prolonged orbits with near-radial velocities,
and come close to the system's center.
The stability of such systems remains an open question to this day,
although several elements suggest out that spherically symmetric
systems of radial orbits could be unstable, and susceptible to lose
their initial sphericity. This is called the radial orbit instability,
henceforth referred to as \roi.
This mechanism could actually be an explanation for the shape of
some astrophysical objects, such as elliptical galaxies, that are hard to
explain otherwise.

For sufficiently large values of $N$, which is certainly the case for galaxies,
$N$-body self-gravitating systems can be described by a smooth distribution
function, and two-body interactions or ``collisions'' can be neglected
in front of the whole gravitational potential created by this function.%
\footnote{To see that, we can consider the two characteristic time constants
of its dynamics.
\emph{Crossing time} is the typical time a particle takes to go across
the system during its movement, it describes its trajectory in the
potential created by other particles as a whole.
\emph{Relaxation time} is the time it takes for a particle's trajectory
to be significantly influenced by collisions.
A calculation, based on the velocity change during an encounter
and the typical number of encounters during one crossing, shows
(see for instance \citet{binntre}, chapter 1.2.1.)
that relaxation time is typically of order $\frac{N}{\ln N}$ larger than
crossing time.
This means that for durations comparable to the crossing time,
the effect of collisions can legitimately be neglected.}
For simplicity, we will assume that each body has the same mass $m$.

Combining the Vlasov equation
(describing how a distribution function evol\-ves in time in a given potential
when collisions are neglected) and the Poisson equation
(giving the potential created by this very function),
we get the Vlasov--Poisson system~:
\begin{equation}
	\left\{ \begin{array}[c]{l}
		\frac{\partial f}{\partial t}
		+ \frac{\mathbf{p}}{m} \cdot \nabla_{\mathbf{q}} f
		- m \nabla_{\mathbf{q}} \psi \cdot \nabla_{\mathbf{p}} f
	= \frac{\partial f}{\partial t} + \left\{ f, E \right\}  = 0\\
	\psi(\mathbf{q}) =
	- G m \int \frac{f'}{|\mathbf{q}-\mathbf{q'}|} \mathrm{d} \Gamma'
	\end{array}
	\right.
\end{equation}
Here, $\mathrm{d} \Gamma$ is a shorthand for
$\mathrm{d}^3 \mathbf{q} \mathrm{d}^3 \mathbf{p}$, and
the notation $a'$ does not denote a derivative of $a$, but
instead the value $a(\mathbf{q'}, \mathbf{p'}, t)$ (for any variable $a$).
$E$ is the average energy per particle
$E\left( \mathbf{q}, \mathbf{p}, t \right)
:= \frac{\mathbf{p}^{2}}{2m} + m \psi(\mathbf{q})$
(or one-particle Hamiltonian).

Spherically symmetric, stationary solutions to this system can be either
isotropic in velocity space, in which case the distribution function can be
written as a function of the energy per particle alone, $f = f_0(E)$;
or, in the case where velocity distribution is not isotropic,
as a function of energy and angular momentum $f_0(E,L^2)$
\citep[see for example][]{perez_et_al}.%
\footnote{Conversely, a distribution function $f_0(E)$ describes indeed
an isotropic, spherically symmetric equilibrium, and $f_0(E,L^2)$ a
spherically symmetric, anisotropic equilibrium.}

Considering the amount of work done about radial orbit instability (\roi)
over the years, and the importance of this concept, we felt it was necessary
to do a complete timeline of publications about it; this will be the second
part of the present paper.
In a third part, we will present the symplectic method that we use to study
the stability of self-gravitating systems, and its prospects on the study
of radial orbit instability.


\section{Historical overview of radial orbit instability}

\subsection{The pioneers}

The first important work published about \roi{} is an analytical result by
\citet{antonov}. It establishes a differential system for a given
``displacement'' of orbits and the corresponding Poisson equation,
in the limit case of radial orbits.
Instability of the system is then proved by constructing a strict Lyapunov
function for this system, although the proof is unclear.

The same year, \citet{henon} published one of the first numerical
simulations of the problem, using $N = 1000$ spherical shells.
Isotropic, polytropic models $f(E) \propto E^n$ were found to be stable
(as it was obtained by Antonov in the 60s), 
whereas anisotropic systems, for generalized polytropes
$f(E) \propto E^n L^{2m}$,
were found to become unstable when $m \to -1$, which corresponds to a system
with more and more radial orbits.
The method does not allow to see the effect on the position space
(the famous ``bar'').
This article made no reference to the work of \citeauthor{antonov}.

A contrary result was published by the French team of \citet{waterbag}.
Using water bag methods (decomposing distribution functions
as a sum of functions constant over phase space domains),%
\footnote{Five years earlier, Doremus, Feix and Baumann had
obtained the stability of isotropic systems by the same method.}
they predicted that all self-gravitating systems $f(E,L^2)$
were stable against non spherical perturbations.


\subsection{Subsequent advances}

In the eighties, \citet{polyach}
proposed a matrix formulation of the stability problem, based on a Fourier
series decomposition of perturbations;
this allows them to prove that models of the form
$f(E - \frac{\lambda L^2}{r_a^2})$
(later called Ossipkov--Merritt models)
are unstable if $r_a^2$ is sufficiently small.
The article unfortunately isn't very clear either,
but the result is contrary to \citeauthor{waterbag} and confirms Antonov and
Hénon.
This article also presents an often-quoted stability criterion,
stating that radial orbit instability occurs when
$\frac{2T_r}{T_\perp} > 1.75 \pm 0.25$,
$T_r$ and $T_\perp$ being respectively the system's total radial and
orthogonal kinetic energies.

The first full, realistic numerical survey of the problem of
gravitational collapse was realized by \citet{albada}.
This survey considers sets of $N=5000$ particles, for different initial
conditions~:
homogeneous spheres and systems made of smaller homogeneous spheres (clumps),
for different speed distributions determined by the initial virial ratio.
The results are clear~: while collapsing homogeneous spheres are not affected by
radial orbit instability,
clumped systems with violent collapse (small virial ratio)
lead to a triaxial equilibrium.
The resulting light and density profiles are compatible
with the ones observed on galaxies when projected onto the sky : the de Vaucouleurs $R^{1/4}$ where $R$ is the projected radius.

One of the first complete studies of instability, both analytical and
numerical, was made by \citet{barneshut}.
Numerical studies with $N$-body methods confirm and complete Hénon's results
for generalized Plummer models, as well as the Russians' stability
criterion.
Their analytical explanation for \roi{}
links it to Jeans instability, stellar pressure in the
tangential direction would no longer be sufficient to offset the natural
tendency of radial orbits to condense.

The next paper on the subject%
\footnote{It actually references \citet{barneshut}, in spite of an apparently
earlier publication date.}
by \citet{merritt_aguilar}, focuses on numerical results.
It uses $N$-body simulations with $N = 5\cdot 10^3$,
taking as initial conditions ``galactic type'' density profiles.
Those models are Ossipkov--Merritt models
(which are isotropic near the center and anisotropic towards the borders)
generated from the well-known Jaffe model
$\rho(r) \propto (r/r_0)^{-2}(1+r/r_0)^{-2}$
(compatible with the projected $R^{1/4}$ profile).
The results are that transition between stability and
instability is fairly sharp, and happens for
$\frac{2 T_r}{T_\perp} \approx 2.5$,
a bit more than predicted by the Russian criterion.
However, comparison with a distribution function decreasing in $E$ and in $L^2$
seems to point out that the value of $\frac{2 T_r}{T_\perp}$
isn't a reliable stability criterion.
The article also puts forward (apparently for the first time)
the idea that \roi{} can be useful in explaining galactic formation.


Linear stability of isotropic spheres against all perturbations, and
of anisotropic spheres against radial ones was achieved by
\citet{Kandrupetsygnet} using energy methods based on an old principle by
\citet{Kulsrudetmark};
this powerful and simple method requires the construction of a Hermitian
operator.
This technique seems irrelevant for the study of radial orbit
instability, which is naturally associated to an anisotropic system receiving
non-radial linear perturbations, for in their case the operator is no longer
Hermitian.
However, a global or energetic approach of this problem was proposed by
a short original but nebulous paper by \citet{maybinney}. When it is applied
to study the stability against non-radial perturbations of a family of
anisotropic isochrone spheres, \citet{maybinney} claim that it is compatible
with the Russian criterion by Polyachenko and Shukhman. Two year later, a
technical paper by  \citet{goodman} clarifies the method proposed by May
and Binney, but no application to the radial orbit instability is proposed.


The English team of \citet{palmerpapa} made a purely analytical study
(the first one since the alleged Russian result,
disregarding the waterbag result of \citeauthor{waterbag})
based on a spectral analysis of perturbations, decomposing them once again
on a family of orthogonal functions. This study seems to indicate
an instability, although is seems very hard to verify.
Two other important aspects of this paper are a ``demonstration'' that the
Russian criterion (already shaken by Merritt and Aguilar) is invalid, and a
explanation of a mechanism for instability growth, inspired by a work
by \citet{lyndenbell}~: an bar-like perturbation of the potential
in a spiral galaxy could influence a star's orbit and lengthen it,
which tends to align orbits along the perturbation.
This effect could play a part in bar formation in spiral galaxies.

A clear synthesis of all those results was made by \citet{merritt1987}, 
including Lynden-Bell's mechanism, as well as a criticism of Jeans instability
mechanism for needing an homogeneous system which is not the case here.

\citet{katz} brought in a new kind of simulations in this context, which
showed that ``authentic'' cosmological simulations with a Hubble flow and
merging phases tend to erase traces of possible primordial \roi.
The same year, an article by \citet{saha} extended the reach of spectral
methods for normal modes to infinite-extension models, which was not the
case for previous studies.
Still the same year, a study by \citet{weinberg} used the matrix methods
initiated by the Russian school of Polyachenko, and found some results again.
This analysis was later the subject of a complete
article by the Argentine team of \citet{cincotta}, who studied transformation
of loop orbits into bow orbits, in the style of Lynden-Bell's mechanism and
in accordance to the intuition of \citet{merritt1987}.


\subsection{A renewed interest}

Taking advantage of some of their analytical results, \citet{perez_et_al}
proposed and tested a new stability criterion for self-gravitating systems
based on the nature of the perturbations it is submitted to.
This criterion is validated on Ossipkov--Merritt models applied to polytropes,
the number of particles involved reaching for the first time reasonable values
of $N \approx 10^4$ for the whole of simulations. In their analytical results,
they explain how waterbag methods are lacking in the field of radial orbits,
which may explain the now-discredited result of \citet{waterbag}.

The German team of \citet{theis} undertook an extensive numerical study of
\roi, using dedicated ``\textsc{Grape}'' machines for
collapses of Plummer spheres with varying initial temperature.
Growth rate of \roi{} is largely affected by potential softening, and very
little by variations of the number of particles.
Those simulations highlighted a very long-term evolution (its time scale is
relaxation time) of triaxial systems produced by \roi{} towards a
more or less spherical system, an evolution that according to the authors is
due to collisions.

A systematic study of gravitational collapse with tests for several numerical
parameters ($N$, softening, \ldots) by \citet{roy} allowed, among other
results, to stress that \roi{} is dependent on the presence of robust
inhomogeneities in the precollapse system. Only several-scale collapses can
lead to \roi~: collapses of homogeneous spheres fail to do so.
Those results were completed and refined by \citet{boily}, who showed a
small effect of particle number on the final state.

Although the role of \roi{} in structure formation was hurt by the
aforementioned work of \citet{katz}, complementary analyses by the
German team  of \citet{huss} and by the Canadian team of \citet{macmillan}
observed the result of medium-scale structure formation by collapse
experiments, with the possibility of numerically suppressing \roi.
Acceptable density profiles are only found when \roi{} actually takes place during
primordial phases; otherwise structure profiles are incompatible with
simulations and observations! This gives a new argument for \roi{}
as a fundamental process of structure formation.

A new activity in this domain is rising since 2005, notably from
E.~Barnes's team.
Their articles, notably \citet{barnes2005}
and \citet{ROI_Moderne}, show that \roi{} not only creates a triaxial system
in position space (which was known for a long time), but also creates a
spatial segregation in velocity space (with an isotropic center and a radial
halo). This segregation could be the root of the universal profile observed
in large structures.
Follow-up papers, including \citet{barneslanzel}, indicated that \roi{}
is not found in constant-density collapse, a statement backed up
by the Italian team of \citet{trenti}.
This agrees with the previous explanation by \citet{roy}, that \roi{}
does not happen without a primordial equilibrium state caused by a several-scale inhomogeneous collapse.

Lastly, \roi{} has been observed in a triaxial state by \citet{antonini}~:
it apparently happens when this state is populated with too many
``box orbits with predominantly radial motions''.
The system would then become more prolate, and still triaxial.

As this overview shows, \roi{} is a phenomenon that has been known for almost
40 years, but still sparks some controversy and contradictory results.
It seems to be fundamental in structure formation, 
yet there are still a lot of unanswered questions
on both the physical and the analytical sense.
Under which conditions does it happen, if it happens?
Can we explain its mechanism?
Several tools can be used to study radial orbit systems,
and among them, potential energy methods could
prove quite useful. We chose to focus on the symplectic method,
which will be covered in the next section.


\section{The symplectic method}

\subsection{Presentation}

The symplectic method is a way of investigating the stability of a
steady state against possible perturbations.
It was first developed by \citet{bartho},
though its use in the study of gravitational plasmas is a relatively recent
development. See for instance \citet{Mor111}, \citet{Mor222} and \citet{kandrupstability}.

This method makes use of the Hamiltonian structure of the system under
scrutiny. The Vlasov--Poisson system indeed derives from the following
Hamiltonian~:
\[
	H \left[ f \right]  =
	\int \mathrm{d} {\Gamma}
		\frac{\mathbf{p}^{2}}{2m} f \left( {\Gamma},t \right)
	- \frac{1}{2} Gm^2 \int \mathrm{d} {\Gamma} \int\mathrm{d}{\Gamma}^{\prime}
		\frac{f \left({\Gamma},t \right) f\left( {\Gamma}^{\prime},t\right)}%
		{\left\vert \mathbf{q} - \mathbf{q}^{\prime} \right\vert }
\]
which is just the system's total energy. Note that its functional derivative
$\frac{\delta H}{\delta f}$ is the one-particle Hamiltonian
$E = \frac{\mathbf{p}^{2}}{2m} + m \psi$.

This structure allows us to easily compute the time variation of any functional
of the distribution function. If $K[f]$ is a derivable functional of $f$, then
by definition of $\frac{\delta K}{\delta f}$~:
\begin{equation}
	\frac{\mathrm{d} K[f]}{\mathrm{d} t}
	= \int \frac{\delta K}{\delta f}
		\frac{\partial f}{\partial t} \mathrm{d} \Gamma
	= \int \frac{\delta K}{\delta f} \left\{ E, f \right\} \mathrm{d} \Gamma
	\label{derivk}
\end{equation}
Here, we use the noncanonical Poisson bracket, which was introduced by
\citet{morrison} for studies in hydrodynamics and magnetohydrodynamics
(and thus sometimes referred to as the Morrison bracket).
For two functionals $A$ and $B$ of $f$, it is defined by
\[
	\left[ A, B \right](f) :=
	\int f \left\{
		\frac{\delta A}{\delta f}, \frac{\delta B}{\delta f}
	\right\} \mathrm{d} \Gamma
\]
Then, from~(\ref{derivk}) it follows that
\begin{equation}
	\frac{\mathrm{d} K[f]}{\mathrm{d} t}
	= - \int f \left\{
		\frac{\delta H}{\delta f}, \frac{\delta K}{\delta f}
	\right\} \mathrm{d} \Gamma
	= \left[K, H \right](f)
\end{equation}

As shown by \citet{kandrupstability},
all physical perturbations $f^{(1)}$ that $f_0$ can receive
may be written in the form~:
\[
	f^{(1)}\left(  {\Gamma},t \right) = -\left\{ g,f_{0}\right\}
\]
$g$ is called the \emph{generator} of the perturbation.
We denote as $G$ the following operator~:
\[
	G[f] := \int f g \mathrm{d} \Gamma
\]

Writing perturbations in this form allows us to compute the energy variation.
A first-order calculation shows that
\begin{equation*}
	H^{(1)} [f_0] = \left[G, H \right](f_0)
	= - \int g \{ f_0, E \} \mathrm{d} \Gamma
	= 0
\end{equation*}
Since $f_0$ is a steady state, $\{ f_0, E \} = 0$ so the energy variation is
zero at first order. This corresponds to the classic definition of an
equilibrium.

The second-order variation is
\begin{equation*}
	H^{(2)} [f_{0}] = \left[G, [G,H] \right](f_0)
\end{equation*}
The complete calculation of $H^{(2)}$ is then possible.
\begin{equation*}
	[G,H](f) = \int f \{ g,E\} \mathrm{d} \Gamma
\end{equation*}
>From the expression of $E$, we get directly
$\frac{\delta E'}{\delta f} = - \frac{G m^2}{|\mathbf{q} - \mathbf{q'}|}$.

It leads to
\begin{eqnarray}
	\frac{\delta[G,H]}{\delta f}  & = &
	\{g,E\} + \int f' \left\{ 
		g', - \frac{G m^2}{|\mathbf{q} - \mathbf{q'}|}
	\right\} \mathrm{d} \Gamma' \nonumber \\
	& = & \{g,E\} 
	+ \int \frac{G m^2}{|\mathbf{q} - \mathbf{q'}|} \{g',f' \} \mathrm{d} \Gamma'
\end{eqnarray}
>From this, computing $H^{(2)}$ is easy~:
\begin{eqnarray}
	H^{(2)}[f_{0}]
	& = & - \int \left.\frac{\delta [G,H]}{\delta f}\right|_{f=f_{0}} \{g,f_{0}\} \mathrm{d} \Gamma
	\nonumber \\
	& = & - \int \left(
		\{g,E\} + \int \frac{Gm^2}{|\mathbf{q}-\mathbf{q'}|}
		\{g',f'_{0}\} \mathrm{d} \Gamma'
	\right) \{g,f_{0}\} \mathrm{d} \Gamma
	\nonumber \\
	& = & - \int \{g,E\} \{g,f_{0}\} \mathrm{d} \Gamma
	- G m^2 \int\!\!\!\int \frac{\{g,f_{0}\}\{g',f'_{0}\}}{|\mathbf{q} - \mathbf{q'}|}
	\mathrm{d} \Gamma \mathrm{d} \Gamma'
\end{eqnarray}

This method is a very powerful one, since it allows us to obtain $H^{(2)}$
efficiently in a very general case.
Most methods to compute energy variation due to a perturbation
require the knowledge of
the system's actual geometry, that is to say $f_0(\mathbf{r},\mathbf{v})$
and $\psi(r)$, which are difficult to obtain knowing only $f_0(E,L^2)$
(except in some particular cases).
With this method, we do not need to know the system's geometry to compute
the energy variation.


\subsection{Stability criterion}

In the classic case of one particle influenced only by conservative forces,
the stability of an equilibrium is directly linked
to the sign of the second-order energy variation~: if it is positive
semi-definite, then the equilibrium is stable;
otherwise it is unstable.
In the previous part, we derived the second-order energy variation around
an equilibrium state~: its sign should provide a criterion to know whether it
is stable or unstable \citep[as seen, again, in][]{Mor111,Mor222}.
Unfortunately, in more general cases, this is not so simple.%
\footnote{An example is the case of a
charged particle in a negative harmonic potential, with a strong enough
magnetic field. This example is covered in more details in another
paper \citep{future}.}

In the case where $H^{(2)} > 0$ for all generators $g$,
then there is a definitive result by \citet{bartho},
which proves that the system is \emph{stable}.
The symplectic method at least allows to prove the stability of an equilibrium.
However, in the case where there are generators $g$ such that $H^{(2)} < 0$,
there is no definite proof of instability, at least not without more hypotheses.

An important mathematical result was given by \citet{blochmarsden}, in a
quite general case.
Consider a Hamiltonian system with finite dimension, that is
initially at equilibrium. Then we suppose there exists a negative energy mode.
In this case, \citeauthor{blochmarsden} proved that with the addition
of dissipation, the equilibrium becomes spectrally unstable,
from which follow linear and nonlinear instability. 
This kind of instability can be called a \emph{dissipation-induced instability}.

As the Vlasov--Poisson is infinite-dimensional, this result doesn't directly
apply to our problem. More recent works by \citet{krechet} on the
infinite dimensional case seem to indicate that it works in the same
way; there is no general proof yet, but a result seems likely in the
near future.

The previous method allows us to retrieve previous results much
more easily.
In the isotropic case $f_0(E)$, this method proves stability against
all perturbations provided that $\partial_E f_0 < 0$.
In the anisotropic case $f_0(E,L^2)$, it proves (with the same condition
$\partial_E f_0 < 0$ ) stability against all
so-called preserving perturbations, \emph{i.e.} perturbations that verify
 $\{g,L^2\} = 0$, see \citet{perezaly}.

In this context, rigorous and complementary results were proposed by a German
team with very different methods, and are compiled in \citet{reinetguo}.


\subsection{Perspectives for radial orbits}

A system with nearly radial orbits can be easily described with the previous
formalism, as $f_0(E,L^2) = \varphi(E) \delta(L^2)$ with
$\delta$ a function that selects values near $0$ (for instance a Dirac
distribution).
In this case, it is possible to show that for a sufficiently selective
function, there are negative energy modes.
The details have been published in another article
\citep[published online]{future}.

The idea is that $H^{(2)}$ has two main components~:
\[
	H^{(2)} =
	\underbrace{- \int \{g,E\} \{g,f\} \mathrm{d} \Gamma}_{(A)}
	\underbrace{- G m^2 %
		\int\!\!\!\int \frac{\{g,f\}\{g',f'\}}{|\mathbf{q} - \mathbf{q'}|}
		\mathrm{d} \Gamma \mathrm{d} \Gamma'}_{(B)}
\]
The term $(A)$ mostly corresponds to kinetic energy variation,
while $(B)$ corresponds to potential energy variation.
For a distribution function that is sufficiently radial, it is possible to
show that there is a class of generators $g$ breaking spherical symmetry,
such that $(A)$ is negligible in front of $(B)$.
The latter is negative, as it can readily be written as the integral of a
function against its own Laplacian.
This proves the existence of negative energy modes.


\section{Conclusion}

We have shown a criterion to discuss the stability of Hamiltonian systems,
and used it in the case of self-gravitating systems such as star clusters or
galaxies, thanks to the the fact that Vlasov--Poisson is a Hamiltonian system.
This criterion works on initial equilibrium states. The addition of dissipation
seems to imply an instability if a steady state has negative energy modes.
Apparently, systems populated with radial orbits have negative energy modes,
which would mean they are unstable and susceptible to lose their spherical
symmetry, triggering radial orbit instability.

It may be possible to give a physical interpretation of this result. Radial
orbits have no tangential velocity, so they do not precess around the
system center; instead they are confined on a line. Bringing the orbits closer 
is be possible, as they do not precess, and it leads to a lower energy state
as the stars would be on average closer to each other than before.
If a direction had a higher than average density, other orbits would tend
to align in this direction, in a lower energy state; those orbits would
bring on others, and so on, leading to an instability. This process would not
be possible if the stars cannot dissipate energy; hence the necessity of
dissipation to insure that radial orbit instability takes place.

\nocite{*}

\bibliographystyle{plainnat}
\bibliography{report}

\end{document}